# Cryogenic interface-state filling and tunneling mechanisms in strained Ge/SiGe heterostructures

Jingrui Ma[1,2,§], Yuan Kang[1,2,3,§], Rui Wu[1,2], Zheng Liu[1,2], Zong-Hu Li[1,2], Tian-Yue Hao[1,2], Zhen-Zhen Kong[3,4], Gui-Lei Wang[3,4,†], Yong-Qiang Xu[1,2], Ran-Ran Cai[1,2], Bao-Chuan Wang[1,2,3], Hai-Ou Li[1,2,3], Gang Cao[1,2,3,*], and Guo-Ping Guo[1,2,3,5]

[1]*Laboratory of Quantum Information, University of Science and Technology of China, Hefei, Anhui 230026, China*
[2]*CAS Center for Excellence in Quantum Information and Quantum Physics, University of Science and Technology of China, Hefei, Anhui 230026, China*
[3]*Hefei National Laboratory, Hefei, Anhui 230088, China*
[4]*Beijing Superstring Academy of Memory Technology, Beijing 100176, China*
[5]*Origin Quantum Computing Company Limited, Hefei, Anhui 230088, China*

---

\* Contact author: gcao@ustc.edu.cn
† Contact author: guilei.wang@bjsamt.org.cn
§ The authors have equally contributed in this article.

**ABSTRACT**. Traps at the semiconductor-oxide interface are considered as a major source of instability in strained Ge/SiGe quantum devices, yet the quantified study of their cryogenic behavior remains limited. In this work, we investigate interface-state trapping using Hall-bar field-effect transistors fabricated on strained Ge/SiGe heterostructures. Combining transport measurements with long-term stabilization and Schrödinger-Poisson modelling, we reconstruct the gradual filling process of interface states at cryogenic condition. Using the calculated valence band profiles, we further evaluate the tunneling current density between the quantum well and the semiconductor-oxide interface. Our calculation demonstrates that the total tunneling current is consistent with a crossover from trap-assisted-tunneling-dominated transport to Fowler-Nordheim-tunneling-dominated transport under different gate bias regimes. These results refine the conventional Fowler-Nordheim-based picture of interface trapping in strained Ge/SiGe heterostructures and provide guidelines for improving Ge-based quantum device performance by improving barrier crystalline qualities and reducing dislocation-related trap densities.

## I. INTRODUCTION.

Semiconductor spin qubits have emerged as a platform for scalable quantum computing. Among the available material platforms, germanium has attracted significant attention, owing to its strong spin-orbit coupling, weak hyperfine effect and strain-induced splitting between heavy- and light-hole states [1–6]. In recent years, impressive progress has been achieved with germanium hole qubits, such as two-dimensional qubit arrays [7–10], high-fidelity quantum gate [4,11,12], coupling with microwave resonators [13–16] and spin shuttling [8,17]. These advances highlight strained Ge as a promising material for future quantum technologies. However, the practical performance of germanium-based quantum device is strongly limited by interface traps. These traps arise from a high density of dangling bonds at semiconductor-oxide interface due to incomplete oxidation under ambient conditions [18–20] and can capture carriers. Specifically, when a negative gate voltage is applied to a germanium quantum device, the two-dimensional hole gas (2DHG) is accumulated in the Ge quantum well (QW), forming a conduction channel. In this process, free holes in the 2DHG may tunnel through the SiGe barrier and reach the semiconductor-oxide interface, where they are captured by traps and form positive fixed charges [21–23]. The presence of these positive charges partially shields the gate voltage, reducing the effective electric field applied to the quantum well. As a result, the channel's conductivity is weakened and the device drifts away from its intended operating point. In order to compensate this effect, a more negative voltage is then required. However, as long as the interface traps are not fully filled, further trapping and threshold voltage drift will occur, causing long-term instability [6,21–29] and complicating measurements and operation in Ge-based devices.

Despite the critical impact on device stability, interface traps in germanium heterostructures have only been studied in a few specific contexts. Previous studies have qualitatively demonstrated the presence of interface states and their influence on device operation [21,22,30,31], and several works have focused on reducing interface state density by chemical treatments and passivation techniques [20,23,27,32]. However, a deeper understanding of interface-state behavior at cryogenic temperatures remains limited.

In this work, we address this challenge by combining cryogenic transport measurements on strained Ge (sGe) heterostructures with numerical simulations to investigate the behavior and mechanism of interface states. By analyzing the gate-voltage dependence of trapped charge from an inverted Schrödinger-Poisson model, we infer the density of occupied interface states as a function of gate voltage and provide quantitative estimates of traps at low temperature. In addition, we also investigate the microscopic mechanism that supports interface trapping by evaluating tunneling current density within the Ge/SiGe heterostructure. While previous works interpret trapping process in terms of Fowler-Nordheim (F-N) tunneling [21–23], our analysis additionally suggests possible contribution from trap-assisted tunneling (TAT) under certain conditions. This approach that combines experimental and inverse numerical modeling enables us to extract the properties of interface states under cryogenic conditions. Our results not only provide insight into the cryogenic mechanisms governing trap dynamics in strained Ge, but also suggest promising optimization directions in future quantum devices.

## II. EXPERIMENT SETUP

The device under test is fabricated on a reverse-graded strained germanium heterostructure whose profile is illustrated in Fig. 1(a). The Ge/SiGe heterostructure consists of a strained germanium quantum well with thickness of 16 nm and a 32 nm $Si_{0.2}Ge_{0.8}$ barrier layer. This configuration of germanium quantum well is similar to that used in Refs. [5] and [23].

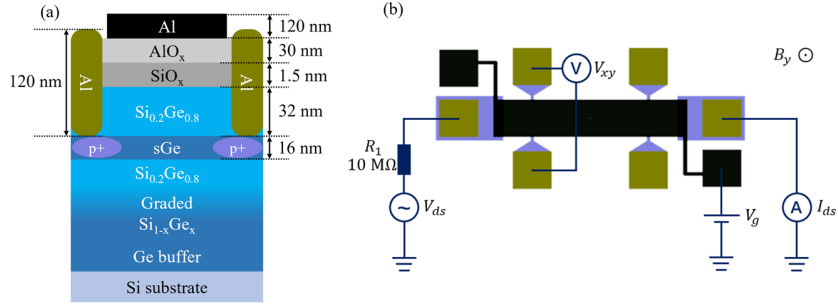

FIG. 1. (a) Cross-sectional view of Hall-bar FET gate stack as well as sGe quantum well configuration. (b) Schematic of FET device and measurement circuit. A resistor $R_1$ is introduced to prevent current overload within the circuit.

In order to observe various phenomena caused by interface defects, the Hall-bar-type field-effect transistor (Hall-bar FET) is fabricated on top of this material stack. As shown in Fig. 1(a) and (b), the enhancement p-type Hall-bar FET consists of a metallic top gate (black), gate oxide (grey), source-drain electrodes (brown) and p-doped contact regions (light purple). When a negative voltage $V_g$ is applied to the top gate, the undoped sGe/SiGe quantum well experiences an electric field across the gate oxide and an accumulation channel will be formed in the QW.

During measurement, the device is mounted in a refrigerator and cooled down to $\approx 4.5$ K. The measurement setup is schematically shown in Fig. 1(b). A DC voltage source is connected to the aluminum top gate to provide gate bias $V_g$. A small AC source-drain excitation $V_{ds}$ is applied using a lock-in amplifier and the resulting current $I_{ds}$ is measured. Meanwhile, an out-of-plane magnetic field $B_y$ is provided by the vector magnet, under which a transverse Hall voltage $V_{xy}$ will be induced perpendicular to the direction of $I_{ds}$ and detected by a second lock-in amplifier.

### III. RESULTS

#### A. Quasistatic hole density measurements

The experiments on source-drain current and hole density are conducted cyclically while sweeping the voltage on top gate, following the protocol illustrated in Fig. 2(a). Each individual cycle is defined by a chosen value of $V_{min}$. At the beginning of each cycle, we firstly set $V_g = 0$ to initialize the measurement. After initialization, $V_g$ is set to $V_{min}$ to turn on the FET. Once 2DHG is formed in the channel, we hold $V_g = V_{min}$ for a duration $T_{meas} \approx 24000$ s, during which $I_{ds}$ is continuously recorded, yielding the trace of time-dependent source-drain current $I_{ds}(t)$. After this waiting period, we sweep $B_y$ and measure transverse voltage $V_{xy}$. By fitting linearly into $\rho_{xy} = B_y/ep_{2DHG} + c$, where $e$ is the elementary charge, $\rho_{xy} = V_{xy}/I_{ds}$ is Hall resistivity and $c$ is a small offset, we extract the hole density $p_{2DHG}$ under this specific $V_g$. This completes one cycle and the protocol is repeated with a more negative value of $V_{min}$.

The full set of $I_{ds}(t)$ under different $V_{min}$ is shown in Fig. 2(b). All curves of current in the figure exhibit a decay before reaching a stabilized value. This behavior is attributed to the leakage within the heterostructure: holes in the 2DHG tunnel through the barrier and are subsequently captured by interface states, as also reported in previous works [23] on Ge-based FETs. We assume that after $T_{meas}$ the system is quasistatic, and the measured $p_{2DHG}$ at this point reflects the final trapped-charge configuration under the corresponding $V_g$. Figure 2(c) illustrates an example of $V_{xy}(B_y)$ traces after current stabilization. Using the relationship of $V_{xy}(B_y)$, we extract carrier density under each $V_{min}$ as summarized in Fig. 2(d). These stabilized values of $p_{2DHG}$ are of more interest in this work since they form the key experimental input for numerical modelling, which is demonstrated in the next section.

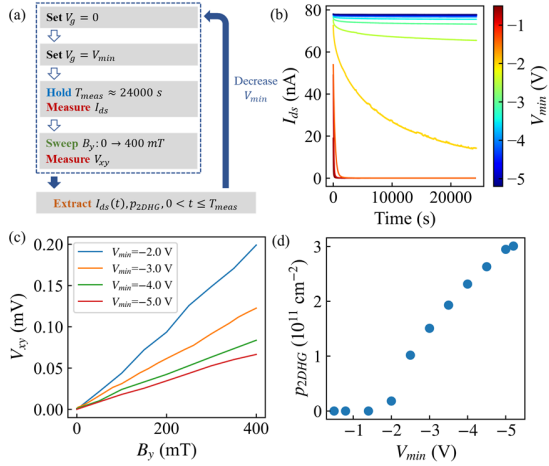

FIG. 2. Quasistatic hole density measurements and results. (a) Measurement protocol of the whole process. (b) Extracted curve of long-term stability of source-drain current $I_{ds}(t)$ under different $V_{min}$. (c) Example of Hall voltage while sweeping the out-of-plane magnetic field. The relationship of $V_{xy}(B_y)$ is then used to compute carrier density in (d). (d) Extracted 2DHG sheet density $p_{2DHG}$ under different $V_{min}$, each point in the figure is extracted after $I_{ds}$ is stabilized, i.e., after the rightmost point of each curve in (b).

### B. Schrödinger-Poisson Modelling and Extraction of Interface Charge

In order to infer the amount of filled interface states from experimental data at each gate voltage, we build an one-dimensional self-consistent Schrödinger-Poisson (S-P) solver [33,34]. The S-P solver takes gate voltage $V_g$ and interface charge density $p_{it}$ as input parameters and returns valence band profile $E_v(x)$ as well as corresponding theoretically calculated carrier density $p_{2DHG,th}$ (see Appendix A for more details). An example of S-P calculation result is shown in Fig. 3(a) for $V_g = -0.5$ V and $p_{it} = 0$. As the valence band edge in Ge QW bends upward and rises above the Fermi level, 2DHG is formed and accumulated in the channel, which is reflected experimentally as a finite $I_{ds}$. In compact form, the S-P simulation can be written as a single mapping:

$$(E_v(x), p_{2DHG,th}) = F(V_g, p_{it}). \quad (1)$$

In our analysis, $V_g$ is given as a boundary condition, the experimental 2DHG density $p_{2DHG,exp}$ is obtained from Hall measurements, and $p_{it}$ is the unknown value to be solved. Thus, extracting $p_{it}$ is equivalent to inverting Eq. (1), which we treat as a root-finding problem. For each fixed $V_g$, we scan different trial values of $p_{it}$, run the S-P solver to get corresponding $p_{2DHG,th}$, and compare it with $p_{2DHG,exp}$. In order to improve numerical stability, we accept the solution when the following condition is satisfied:

$$0.99 \leq p_{2DHG,th}/p_{2DHG,exp} \leq 1.01. \quad (2)$$

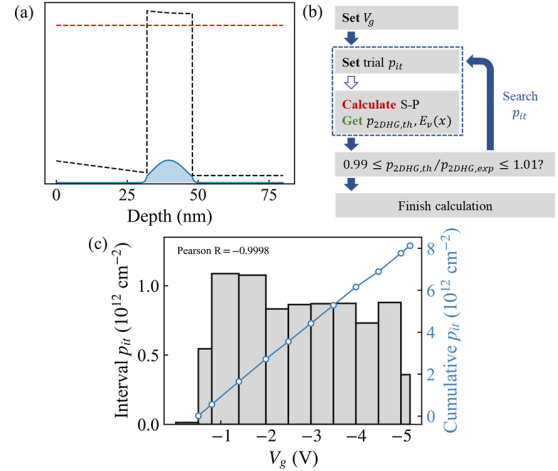

FIG. 3. (a) Example of self-consistent S-P calculation. The black dashed line shows the valence band profile along the 1D mesh, red horizontal line indicates Fermi level. Shaded area shows the simulated hole wavefunction. Note that this profile only shows the SiGe/sGe/SiGe region. (b) Calculation process of inferring $p_{it}$ from $V_g$ and $p_{2DHG,exp}$. (c) Accumulation of $p_{it}$ vs $V_g$ and the amount of interface charge in each gate voltage interval. Pearson R value is calculated using cumulative density and $V_g$.

The $p_{it}$ value that satisfies Eq. (2) is returned as the saturated interface charge under that $V_g$. A summary of this workflow is illustrated in Fig. 3(b). By repeating this procedure for each data point in Fig. 2(d), we would be able to reconstruct how the interface charge is built up with gate voltage. Figure 3(c) illustrates the resulting $p_{it}$ as a function of $V_g$ (blue line), together with the density of interface states being filled in each voltage interval (grey bars). As indicated by Pearson correlation coefficient in Fig. 3(c), we find that the inferred trapped charge increases almost linearly with $|V_g|$ over the explored range, consistent with a capacitor-like response of the semiconductor-oxide interface stack. This inverted S-P method effectively resolves intermediate values of trap filling under cryogenic conditions, which, to our knowledge, have not been quantified in existing studies.

**C. Tunneling current density Calculation**

The analysis in the previous section provided the equilibrium density of occupied interface states as a function of gate voltage. Based on these results, we intend to further study how holes in 2DHG reach interface traps, and investigate the microscopic tunneling process across the quantum well barrier. Specifically, we use the valence band profile obtained from S-P solver (Eq. (1)) as input to calculate the tunneling current density. Our goal is to estimate the tunneling current density from the QW to the semiconductor-oxide interface and identify the tunneling mechanism under the experimentally explored gate-voltage interval.

The tunneling current density is taken as the sum of three distinct tunneling mechanisms: direct tunneling, Fowler-Nordheim (F-N) tunneling and trap-assisted tunneling (TAT) [35], as illustrated in Fig. 4(a). The total current density is written as:

$$J_{total} = J_{direct} + J_{FN} = J_{TAT}. \quad (3)$$

Direct tunneling occurs when holes in Ge QW coherently tunnel through the entire SiGe barrier and reach the interface. For a rectangular barrier, the corresponding current density can be approximated as [36]:

$$J_{direct} \propto \exp\left(-\frac{2t\sqrt{2m^*}\phi_B}{\hbar}\right), \quad (4)$$

where $\phi_B$ is the barrier height, $m^*$ is effective mass and $t$ is the barrier thickness. When the band bending becomes strong (i.e., under large negative gate bias), the barrier is reshaped approximately into a triangle and the tunneling is described by Fowler-Nordheim theory [37]. The F-N tunneling current density through a triangular barrier is given by [35–37]:

$$J_{FN} = \frac{e^3 E^2}{8\pi h \phi_B} \exp\left(-\frac{8\pi\sqrt{2m^*}\phi_B^{3/2}}{3ehE}\right), \quad (5)$$

Where $E$ is the electric field across the barrier. In our calculation, both barrier height $\phi_B$ and electric field $E$ are obtained from S-P simulation results.

Trap-assisted tunneling refers to a multi-step process in which holes reach the semiconductor-oxide interface via intermediate defect states in the barrier. Strain-induced misfit dislocations in the SiGe alloy may act as such intermediate sites and create tunneling pathways. Here, unless otherwise specified, the "trap" discussed in this section refers to defects within the barrier layer, instead of the interface states discussed previously. Existing studies report that typical dislocation densities in SiGe could reach the order of $10^6$ to $10^8$ cm$^{-2}$ [2,38–40], which we consider sufficient to provide a dense net of intermediate sites for hole tunneling. The TAT current density is obtained by integrating the contribution over the whole layer since the process may occur at any position within the barrier. The analytical form of $J_{TAT}$ is written as [35]:

$$J_{TAT} = e \int_0^t \frac{N_T(x)}{\tau(x)} dx, \quad (6)$$

where $N_T(x)$ is the distribution of trap, $\tau(x)$ is a parameter that combines the trap's local capture and emission rate. Numerically, we implement the calculation of $J_{TAT}$ by:

$$J_{TAT} = e \sum_{i=1}^m p_i \cdot \sigma \cdot \left(\sum_j \Delta x \cdot N_{T,j} \cdot TC_j\right), \quad (7)$$

Where $p_i$ is the supply hole density from the i-th subband, $\sigma$ is the hole capture cross section. The supply hole density at each subband is given by:

$$p_i = k_B T \frac{m^* m_e}{\pi \hbar^2} \ln\left[1 + \exp\left(\frac{E_i - E_F}{k_B T}\right)\right], \quad (8)$$

where $E_i$ is the i-th eigenenergy computed from Schrödinger's equation, $E_F$ is the Fermi level, $m_e$ is the electron mass and $T = 4.5$ K is the temperature of simulation. The last term in Eq. (7) is the integration's numerical form. $\Delta x$ is the mesh size, $N_{T,j}$ is the trap density at the j-th point and $TC_j$ is the tunneling coefficient. The tunneling coefficient reflects the rate of hole tunneling between adjacent trap sites, and is evaluated under the Wentzel-Kramers-Brillouin (WKB) approximation [41–43]. The trap density $N_{T,j}$ is sampled from a distribution in both space and energy:

$$N_{T,j} = w(\mu_{trap}, \sigma_{trap}) N_T(x), \quad (9)$$

where $w$ is the weight indicating trap distribution in energy. In this work, we assume that traps follow a Gaussian distribution in energy [35], characterized by a specified mean value $\mu_{trap}$ and standard deviation $\sigma_{trap}$. Meanwhile, $N_T(x)$ is the trap density calculated from a spatial distribution. Dislocation-related traps are considered to originate at the Ge/SiGe interface, so that the trap density is maximal at this interface and decrease as moving away from it. We use an exponential decay profile to phenomenologically describe this distribution along the growth direction:

$$N_T(x) = N_{trap} \exp\left(-\frac{x - x_0}{\lambda}\right), \quad (10)$$

where $N_{trap}$ is the largest trap density at Ge/SiGe interface $(x = x_0)$, and $\lambda$ is a characteristic length of a few nanometers.

Using the equations above, the calculated tunneling current density for direct tunneling, F-N tunneling and TAT are shown in Fig. 4(b), (c), (e) respectively. The current density associated with direct tunneling is constantly negligible (approximately $10^{-30}$ A/m$^2$) over the entire gate-voltage range, which is expected since $J_{direct}$ scales exponentially with thickness of the barrier according to Eq. (4). In our heterostructure, the $t = 32$ nm barrier (Fig. 1(b)) strongly suppresses direct tunneling, which was also suggested by existing works [21,22]. In contrast, the F-N current density experiences a strong turning point at $V_g \approx -4.0$ V. This behavior reflects $J_{FN}$'s strong dependence on electric field, which is directly linked to the valence-band profile. As shown in the insets of Fig. 4(c), the Ge/SiGe valance band evolves from weak to strong bending as more negative $V_g$ is applied. The specific value of this turning point is mainly determined by the thickness of the SiGe barrier, as illustrated in Fig. 4(d). Notably, Figure 4(d) is simulated without considering interface traps, hence the simulation range of voltage differs from that in other figures. For a thicker barrier, holes more difficult to tunnel through, and stronger gate voltage is required to achieve the same tunneling current.

Meanwhile, TAT current density, shown in Fig. 4(e), remains mostly within the same order of magnitude within the explored range of $V_g$. Here, we observe a slight decrease of $J_{TAT}$ at the most negative gate voltages. In our model, this behavior reflects the fact that TAT is primarily governed by trap properties [35]. While the WKB tunneling coefficient $TC$ increases with $|V_g|$ as the electric field is enhanced, this term does not outperform the change in trap distribution $N_T$. Dislocation-related traps in SiGe have been reported at energies above the valence-band edge at approximately 140 meV [44,45], whereas the valence-band offset between sGe QW and SiGe barrier in our heterostructure is only ~114 meV [22,46]. As the gate bias becomes more negative, the stronger band bending makes subband energies move further away from $\mu_{trap}$, hence reducing the Gaussian weight $w$ associated with the total trap density (Eq. (9)) and compensating the increase in tunneling coefficient. We emphasize that this decrease in $J_{TAT}$ does not affect the following analysis of tunneling current contribution. It should also be noted though, since many microscopic trap parameters are not directly accessible in our experiment, the $J_{TAT}$ calculation is intended to provide an order-of-magnitude estimate, rather than to obtain a fairly precise value of $J_{TAT}$. We used dislocation-related trap parameters reported in literatures, as summarized in TABLE I, to perform a parameter sweep and support our TAT estimation (see Appendix B for more details). The grey shaded area in Fig. 4(e) indicates the reasonable range of $J_{TAT}$ corresponding to our device. For our Hall-bar FET with gate area of $S \approx 1 \times 10^{-7}$ m$^2$, the current density in the shaded area provides tunneling current between 10 pA to tens of nA, which is consistent with measured channel current in Fig. 2(b). We observe that under physically sound trap parameters, the results of $J_{TAT}$ mostly fall into this range, hence providing a possible leakage path with the correct order of magnitude to account for the interface-state filling under small $|V_g|$ observed in the experiment.

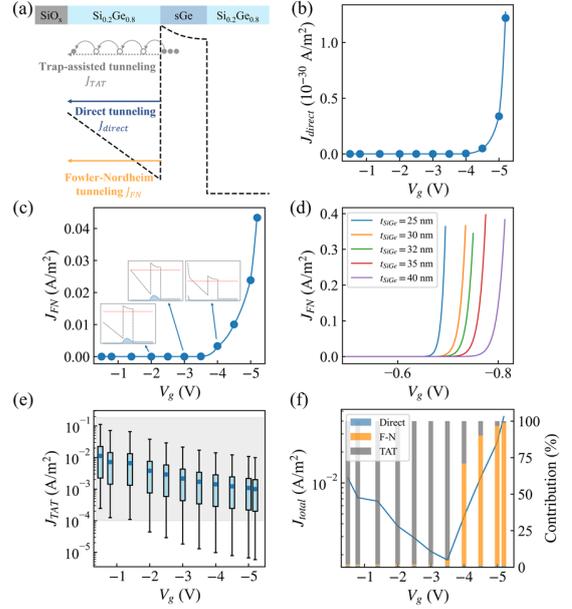

FIG. 4. (a) Illustration of direct tunneling, Fowler-Nordheim tunneling and trap-assisted tunneling within the Ge/SiGe heterostructure. (b) Direct and (c) F-N tunneling current density as a function of $V_g$. Insets: Valence band simulation at $V_g = -2, -3, -4$ V respectively. (d) Comparison of $J_{FN}$ simulated under different SiGe barrier thickness. Note that this calculation is performed assuming $p_{it} = 0$. (e) TAT current density as a function of $V_g$. Blue scatters indicate $J_{TAT}$ under default parameter, boxes indicate $J_{TAT}$ under varied trap parameters in TABLE I, grey shaded area indicates $J_{TAT}$ order consistent with experiments. (f) Total current density (blue line) as well as the contribution percentage of each mechanism. Blue bar indicates direct tunneling (≪1%), orange bar indicates F-N tunneling and grey bar indicates TAT contribution.

TABLE I. Key parameter range used in TAT current density calculation in strained Ge/SiGe heterostructure.

| Parameter | Baseline value | Test range | Unit | Refs. |
|---|---|---|---|---|
| $N_{trap}$ | $1 \times 10^7$ | $[1 \times 10^6, 1 \times 10^8]$ | cm$^{-2}$ | [2,39,40,47] |
| $\sigma$ | $8 \times 10^{-17}$ | $[5 \times 10^{-18}, 2 \times 10^{-16}]$ | cm$^2$ | [45,48] |
| $\mu_{trap}$ | $E_v + 140$ | $[E_v + 120, E_v + 160]$ | meV | [44,45] |

The total tunneling current density, along with the relative contribution of each mechanism, is then plotted in Fig. 4(f). The direct-tunneling contribution remains well below 1% for all gate voltages. It is therefore clear that the tunneling process in strained Ge is effectively a two-mechanism problem, dominated separately by TAT and F-N tunneling in different gate voltages. At small $|V_g|$, the electric field within the heterostructure is insufficient to support significant F-N tunneling, and holes in 2DHG most likely travel through dislocation-induced traps in SiGe barrier towards the surface. In this case, we suggest that the total tunneling current is dominated by TAT under physically reasonable parameters, after which holes are captured by interface states. As the gate bias becomes more negative, although the contribution of TAT still exists, its weight becomes smaller compared to the rapidly increasing F-N tunneling current density supported by electric field. Therefore, under the $V_g \leq -4.0$ V regime, a description based on F-N is a good approximation of the tunneling-trapping process. Comparing to previous studies that qualitatively interpreted interface trapping in terms of F-N tunneling alone [21–23], our calculations indicate that for realistic barrier parameters, an F-N-only model underestimates the current at small and moderate fields. This extends the conventional F-N-based picture by demonstrating that TAT supported by dislocation-related defects provides a natural explanation for the observed interface accumulation at small negative voltages.

## IV. DISCUSSION

To further support our tunneling analysis and test whether the prominence of TAT at small and moderate fields is an artifact of uncertain trap parameters, we perform the same analysis to a representative strained-Si/SiGe quantum well. As shown in Fig. 5(a), the reference structure contains a 15 nm undoped Si layer buried 25 nm deep into the Si$_{0.7}$Ge$_{0.3}$ barrier, with a conduction band offset of approximately 180 meV [46,49,50]. Following the procedure described in Sec. III C, we first use the S-P solver to obtain the self-consistent conduction band profile (Fig. 5(b)), then compute the corresponding $J_{FN}$ and $J_{TAT}$ in silicon using literature-based defect parameters summarized in TABLE II.

The comparisons of $J_{FN}$ and $J_{TAT}$ between Si/SiGe and Ge/SiGe heterostructures are illustrated in Fig. 5(c) and Fig. 5(d) respectively. Since Si- and Ge-based devices have different working regime, we choose to use electric field, rather than gate voltage, to better visualize data. We find that F-N tunneling current densities are of similar magnitude for both stacks, consistent with the fact that $J_{FN}$ is controlled by the electric field. Differences in the specific onset of $J_{FN}$ mainly reflect influence of other barrier parameters (barrier height, effective mass and thickness) and is beyond the scope of this work. By contrast, the calculated $J_{TAT}$ in strained Si is suppressed by roughly one to three orders of magnitude compared to the Ge case under moderate field range. This could be explained naturally by the difference in trap parameters between TABLE I and TABLE II. In particular, the dislocation-related trap density assumed for Si/SiGe is lower, and the trap energy distribution is located far deeper relative to the tunneling window, hence reducing the effective contribution of traps in calculation. Although the effective electron capture cross section in Si is significantly larger than the corresponding hole capture cross section in Ge, this does not outperform the other two factors, yielding a total suppression of $J_{TAT}$ in the Si/SiGe heterostructure.

These comparative results support a physical picture where defect-mediated pathways could enable significant tunneling at small and moderate fields in strained Ge/SiGe, whereas the high-electric-field regime is governed predominantly by F-N tunneling in both material systems. Experimentally, the pronounced current decay observed in our Ge/SiGe devices (Fig. 2(b)) is consistent with this two-regime picture, while related studies on Si/SiGe report slower evolution of channel resistivity [51] in the absence of enough defect-mediated tunneling pathways from our calculation.

TABLE II. Key parameter range used in TAT current density calculation in strained Si/SiGe heterostructure.

| Parameter | Baseline value | Test range | Unit | Refs. |
|---|---|---|---|---|
| $N_{trap}$ | $1\times10^6$ | $[1\times10^5, 1\times10^7]$ | cm$^{-2}$ | [52–55] |
| $\sigma$ | $4\times10^{-12}$ | $[1\times10^{-13}, 1\times10^{-12}]$ | cm$^2$ | [56] |
| $\mu_{trap}$ | $E_c - 600$ | $[E_c - 620, E_c - 580]$ | meV | [56] |

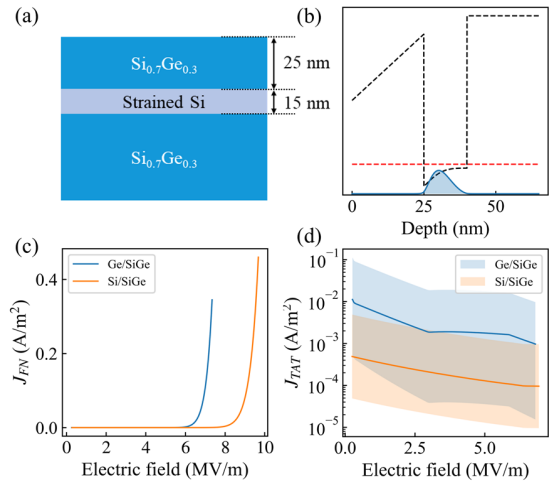

FIG. 5. (a) Representative material stack used in calculation. Note that no interface states are considered. (b) Example of conduction band profile and electron wavefunction from S-P solver. (c) Comparison of $J_{FN}$ and (d) $J_{TAT}$ between Ge/SiGe and Si/SiGe heterostructures. Shaded areas in (d) indicate $J_{TAT}$ estimated under varied trap properties in TABLE I, II for Ge and Si QW respectively.

In addition to supporting the tunneling interpretation, our analysis highlights practical directions for heterostructure optimization. Increasing the barrier thickness can suppress F-N tunneling, as shown in Fig. 4(d). However, under small and moderate field regimes, which are most relevant to qubit operation, suppressing TAT pathways by defects in the SiGe barrier is expected to be particularly important. Improving crystalline quality during heterostructure growth to reduce dislocation-related trap densities should therefore strongly mitigate TAT and enhance device stability. This perspective is highly consistent with recent reports on lattice-matched Ge/SiGe heterostructures grown below the critical thickness [57–59], where significant improvements in material property and qubit performance have been reported [60–63]. Besides lattice matching, several complementary strategies have also been demonstrated. On the materials side, optimized buffer designs and growth protocols can reduce threading-dislocation density in Ge-rich quantum wells [40,64]. On the device side, introducing a backgate provides an additional degree of freedom to tune the location of hole wavefunction [28], potentially reducing its overlap with defect-rich regions. Finally, surface passivation that reduce the semiconductor-oxide interface trap density [23,27,65–68] directly mitigate charge trapping near the surface. Taken together, these engineering approaches provide a practical pathway improving stability and scalability of germanium-based quantum devices.

## V. CONCLUSIONS

In conclusion, we fabricated Hall-bar field-effect-transistors on strained Ge/SiGe heterostructures and investigated the mechanism behind interface-state trapping. Transport measurements after long-term stabilization at cryogenic temperatures as a function of gate bias allow us to extract quasistatic carrier density after the stabilization of transient current. Using these densities as constraints in an inverted Schrödinger-Poisson model, we reconstruct the sheet density of occupied interface states as a function of gate voltage, thereby quantifying the progressive filling of interface traps at low temperature. We then apply the simulated band profiles in tunneling models to evaluate the current density associated with direct, Fowler-Nordheim and trap-assisted tunneling. The combined experimental and modelling results support a two-regime transition from TAT-dominated at small and moderate fields to F-N-dominated tunneling at high fields, which extends the conventional interpretation that attributes interface trapping in strained Ge solely to F-N tunneling.

Beyond establishing this mechanism, our calculation provides practical guidance for device optimization. While increasing barrier thickness suppresses high-field leakage by F-N tunneling, mitigating tunneling pathways from dislocation-related defects is of importance. This highlights the necessity to improve crystalline quality and reduce dislocation trap densities as key routes to enhance future Ge-based quantum devices.

**ACKNOWLEDGMENTS**

This work was supported by the National Natural Science Foundation of China (12574552, 12474490, and 12304560), and Quantum Science and Technology-National Science and Technology Major Project (Grant No. 2021ZD0302300). This work was partially carried out at the University of Science and Technology of China Center for Micro and Nanoscale Research and Fabrication.

**DATA AVAILABILITY**

The data that support the findings of this article are not publicly available. The data are available from the authors upon reasonable request.

**APPENDIX A: SELF-CONSISTENT SCHRÖDINGER-POISSON CALCULATION**

In this section, we describe the details of numerical calculation. The self-consistent Schrödinger-Poisson (S-P) solver involves Schrödinger and Poisson equation coupled in two directions [33]. The 1D Schrödinger equation used in this model is time-independent and described by:

$$-\frac{\hbar^2}{2}\frac{d}{dx}\left(\frac{1}{m^*(x)}\frac{d}{dx}\right)\psi(x) + V(x)\psi(x) = E\psi(x), \quad (A1)$$

where $\psi$ is the wavefunction, $V$ is the potential, $E$ is the eigenenergy and $m^*$ is the effective mass. Solving Eq. (A1) yields a finite number of wavefunctions with corresponding eigenenergies. Meanwhile, the one-dimensional Poisson equation reads:

$$\frac{d}{dx}\left(\epsilon_r(x)\frac{d}{dx}\right)\phi(x) = -\frac{e}{\epsilon_0}p(x), \quad (A2)$$

where $\epsilon_r$ is the relative permittivity, $\phi$ is the electrostatic potential and $p$ is the distribution of holes. On one hand, the resulting $\phi$ in Poisson equation is translated into $V$ in Schrödinger equation by:

$$V(x) = -e\phi(x) + \Delta E_v(x), \quad (A3)$$

where $\Delta E_v$ is the valence band offset between germanium and $Si_{0.2}Ge_{0.8}$ in the QW. On the other hand, the hole wavefunction from Schrödinger is translated into hole density by:

$$p(x) = \sum_{i=1}^{m}\psi_i^*(x)\psi_i(x)p_i(x), \quad (A4)$$

where $m$ is the total number of bound states and $p_i$ is the occupation of holes within each state (subband). The concentration of holes for a certain state is represented by

$$p_i(x) = \frac{m^*(x)}{\pi\hbar^2}\int_{-\infty}^{E_i}\frac{1}{1+e^{(E_F-E)/k_BT}}dE, \quad (A5)$$

where $E_F$ is the Fermi level and $E_i$ is the i-th eigenenergy solved from Eq. (A1).

In order to solve the equations above numerically, we firstly discretize Eq. (A1) along the one-dimensional finite-difference mesh:

$$-\frac{\hbar^2}{2}\frac{1}{\Delta x}\left(\frac{\psi_{i+1}-\psi_i}{m^*_{i+1/2}\Delta x} - \frac{\psi_i-\psi_{i-1}}{m^*_{i-1/2}\Delta x}\right) + V_i\psi_i = \lambda\psi_i, \quad (A6)$$

where $i$ is the index of mesh point, $\Delta x$ is the mesh size of the grid, and $m^*$ with a half-integer index is the mid-point value constructed from effective mass in adjacent points. This reduces a differential equation to an eigenvalue problem of a tridiagonal matrix. Numerically solving Eq. (A6) yields a finite number of wavefunctions with corresponding eigenenergies.

As another part of the iteration, we solve Poisson's equation using Newton's method. Assume in the k-th iteration, we have a guess of electrostatic potential $\phi^{(k)}$, and inside the 1D Poisson equation, we have:

$$\frac{d}{dx}\left(\epsilon_r(x)\frac{d}{dx}\right)\phi^{(k)}(x) = -\frac{e}{\epsilon_0}p\left[\phi^{(k)}\right](x). \quad (A7)$$

Here, we write in $p\left[\phi^{(k)}\right](x)$ to explicitly indicate that the charge density depends on the electrostatic potential profile. This is because the hole density is derived from subbands in Schrödinger's equation, which are again determined by the electrostatic potential from the previous iteration of Poisson (Eqs. (A2), (A4) and (A5)). The left-hand side of Eq. (A7) is simply Laplacian of $\phi^{(k)}(x)$, with additional weights taken from dielectric constants. The right-hand side indicates the charge density calculated from this $\phi^{(k)}$, in other words, the value that the Laplacian should ideally equal to. Eq. (A7) holds only when the iteration is converged and the Poisson potential does not change with further calculations. By taking the difference of two sides, we define the residual function of Poisson equation at iteration $k$:

$$R^{(k)}(x) = \frac{d}{dx}\left(\epsilon_r(x)\frac{d}{dx}\right)\phi^{(k)}(x) + \frac{e}{\epsilon_0}p\left[\phi^{(k)}\right](x). \quad (A8)$$

This equation tells how far we are from satisfying Poisson equation for the current $\phi$. If $R^{(k)} = 0$ everywhere, the self-consistency is then achieved. Using this residual, we would be able to calculate the correction term $\Delta\phi^{(k)}$ by:

$$J\left(\phi^{(k)}\right)\Delta\phi^{(k)}(x) = -R^{(k)}(x), \quad (A9)$$

Where $J\left(\phi^{(k)}\right) = \partial R/\partial\phi$ is the Jacobian matrix, constructed at each iteration numerically. The electrostatic potential for the next iteration is then updated as:

$$\phi^{(k+1)}(x) = \phi^{(k)}(x) + \alpha\cdot\Delta\phi^{(k)}(x), \quad (A10)$$

TABLE III. Key parameters used in the S-P calculation.

| Parameter | Description | Value | Refs. |
|---|---|---|---|
| $E_{\Delta v}$ | Valence band offset | 114 meV | [22,46] |
| $m_{Ge}^*$ | Hole effective mass in Ge | 0.0728 | [5] |
| $t_{SiO_x}$ | SiO$_x$ layer thickness | 1.5 nm | [23] |
| $t_{SiGe}$ | Si$_{0.2}$Ge$_{0.8}$ barrier thickness | 32 nm | [5] |
| $t_{Ge}$ | Ge QW thickness | 16 nm | [5] |
| $\Delta x$ | Mesh size | 0.05 nm | / |
| $\alpha$ | Damping coefficient | [0.0001, 0.8] | / |

with $\alpha$ as a damping coefficient $0 < \alpha \leq 1$ to ensure numerical stability. Note that in Newton's method we did not explicitly solve Poisson's equation as we did in Schrödinger, instead the update value of electrostatic potential in each step of iteration is computed numerically.

The convergence of the calculation is handled by monitoring the maximum change of potential energy, i.e., the valence band profile, after each update:

$$conv = \max \left| V^{(k)}(x) - V^{(k-1)}(x) \right|. \quad (A11)$$

The iteration continues until the criterion $conv < 10$ μeV is satisfied.

While conventional S-P solvers focus on the band profile of semiconductor heterostructures, we extend this by including the dielectric layers to consider the impact of interface charges. Specifically speaking, the material used in calculation is the full AlO$_x$/SiO$_x$/Si$_{0.2}$Ge$_{0.8}$/Ge/Si$_{0.2}$Ge$_{0.8}$ stack (Fig. 1(b)). When solving Schrödinger's equation, we only consider the QW stack of Si$_{0.2}$Ge$_{0.8}$/Ge/Si$_{0.2}$Ge$_{0.8}$ to obtain hole wavefunctions; when solving Poisson's equation, the full stack is used, in which gate voltage $V_g$ is set as a Dirichlet boundary condition from the left side and filled interface traps $p_{it}$ is treated as fixed positive charge within the SiO$_x$ layer. Solving the S-P equation iteratively yields the self-consistent profile of valance band $E_v(x)$ and hole space distribution $p(x)$ under a given $V_g$ and $p_{it}$. In order to transform the space charge density of holes (unit: C/m$^3$) into sheet density consistent with experiments (unit: cm$^{-2}$ or m$^{-2}$), we additionally calculate:

$$p_{2DHG,th} = \frac{1}{e} \int_{x_1}^{x_2} p(x) dx, \quad (A12)$$

where $x_1$ and $x_2$ are indices defining the region of germanium quantum well. This ultimately defines the output of S-P calculation as a single mapping explained in Eq. (1) of main text.

The material parameters used in the S-P solver are obtained from either literatures or linear interpolation

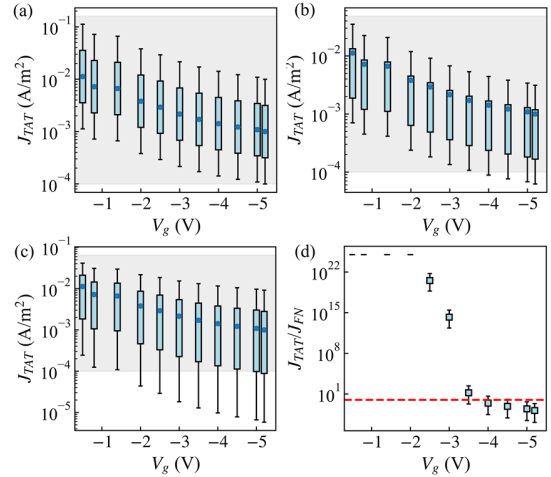

FIG. 6. Sensitivity test results of TAT current density in Ge/SiGe. $J_{TAT}$ is evaluated under varied (a) trap density, (b) hole capture cross section and (c) trap energy respectively. Blue scatters indicate $J_{TAT}$ under default parameter, boxes indicate $J_{TAT}$ under varied trap parameters in TABLE I, grey shaded area indicates $J_{TAT}$ order consistent with experiments. (d) Ratio of $J_{TAT}/J_{FN}$ combining results from (a) to (c). Numerical explosion values with extremely small $J_{FN}$ under small $|V_g|$ are capped and not shown. Red horizontal line indicates reference position where $J_{TAT} = J_{FN}$.

of SiGe alloy. Some of the key parameters are summarized in TABLE III.

**APPENDIX B: SENSITIVITY OF TRAP-ASSISTED TUNNELING CURRENT DENSITY TO TRAP PARAMETERS**

In the main text, trap-assisted tunneling (TAT) current density is calculated using a model that involves several microscopic parameters of dislocation-related traps in SiGe barrier, which are not directly available from our experiment. As a result, the calculated TAT current density $J_{TAT}$ should be regarded as an order-of-magnitude estimate. The purpose of this section is to assess how sensitive the calculations are to the trap parameters and demonstrate that the two-regime picture suggested in the manuscript persists over a physically reasonable parameter range.

We focus on three trap parameters that strongly influence the calculation: the dislocation-related trap density $N_{trap}$, capture cross section for holes $\sigma$ and the mean trap energy $\mu_{trap}$ used in Gaussian energy distribution. For each of the three parameters, we select a range consistent with reported values for dislocation-related traps in Ge or Ge-based platforms. The chosen values are summarized in TABLE I. For each parameter, we set a baseline value and a corresponding test range. While keeping the other two parameters at their baseline values, we recompute $J_{TAT}(V_g)$ using the same S-P valence band profile in the main text, and sweep the parameter under test over the specified range. The results for $J_{TAT}$ are shown in Fig. 7(a), (b) and (c) for variations of trap density, capture cross section and trap energy respectively. Fig. 4(e) in the main text combines all these results. We find that the values of $J_{TAT}$ obtained in parameter tests mostly fall into this range, indicating that TAT provides a possible leakage path with the correct order of magnitude to account for the interface-state filling under small $|V_g|$ observed in the experiment.

In addition, we compare $J_{TAT}(V_g)$ and the Fowler-Nordheim tunneling current density $J_{FN}(V_g)$ for different trap parameters by plotting the ratio of $J_{TAT}/J_{FN}$ in Fig. 7(d). We find that while the absolute $J_{TAT}/J_{FN}$ values depend on trap parameters, the qualitative two-regime picture persists. For all parameter sets in TABLE I, TAT dominates at small and moderate gate bias, whereas F-N tunneling dominates at large negative gate bias, and the crossover between two mechanisms occurs consistently at $V_g \approx -4.0$ V. This confirms that the mechanism discussed in Sec. V and Fig. 4(f) is not from an arbitrary choice of trap parameters, but from values supported by relevant literatures with physical robustness.